 \documentclass[final,5p,times,twocolumn,number]{elsarticle}
\biboptions{sort&compress}

\usepackage{amssymb}
\usepackage{lipsum}
\usepackage{amsthm}
\usepackage{amsmath}

\usepackage[utf8]{inputenc} 
\usepackage[T1]{fontenc} 

\RequirePackage{xcolor}
\RequirePackage[pdfencoding=auto, psdextra]{hyperref}

\journal{Physics Letters B}

\begin{document}

\begin{frontmatter}



\title{ $\Xi_c(2790)^{+/0}$ and $\Xi_c(2815)^{+/0}$ radiative decays}


\author[Tec]{H. Garc{\'i}a-Tecocoatzi \corref{cor1}}
\ead{hugo.garcia.t@tec.mx}
\cortext[cor1]{Corresponding author}
\author[Tec]{A. Ramirez-Morales}
\author[DCI,DIFI,INFN]{A. Rivero-Acosta}
\author[INFN]{E. Santopinto}
\author[CONAHCyT,DCI,DualCP]{C. A. Vaquera-Araujo}

\affiliation[Tec]
{organization={Tecnologico de Monterrey, Escuela de Ingenieria y Ciencias}, 
            addressline={General Ramon Corona 2514},
            city={Zapopan},
            postcode={45138}, 
            country={Mexico}}
\affiliation[DCI]{organization={Departamento de F\'isica, DCI, Campus Le\'on, Universidad de Guanajuato},
            addressline={Loma del Bosque 103, Lomas del Campestre}, 
            city={Le\'on},
            postcode={37150}, 
            state={Guanajuato},
            country={Mexico}}
\affiliation[DIFI]{organization={Dipartimento di Fisica, Universit\`a di Genova},
            addressline={Via Dodecaneso 33}, 
            postcode={16146}, 
            state={Genova},
            country={Italy}}
\affiliation[INFN]{organization={INFN, Sezione di Genova},
            addressline={Via Dodecaneso 33}, 
            postcode={16146}, 
            state={Genova},
            country={Italy}}
\affiliation[CONAHCyT]{organization={Consejo Nacional de Humanidades, Ciencias y Tecnolog\'ias},
            addressline={Av. Insurgentes Sur 1582. Colonia Cr\'edito Constructor, Del. Benito Ju\'arez}, 
            postcode={03940}, 
            state={Ciudad de M\'exico},
            country={Mexico}}
\affiliation[DualCP]{organization={Dual CP Institute of High Energy Physics},
            postcode={28045}, 
            state={Colima},
            country={Mexico}}
            

\begin{abstract}
In this work, we study the $\Xi_c$ baryon electromagnetic decay widths within the constituent quark model formalism through an analysis of the transitions from $P$-wave states to ground states.  We use the non-relativistic limit of the Hamiltonian for the electromagnetic interaction on keeping all the terms up to the order $m_j^{-1}$ and by means of a new algebraic methodology based on ladder operators. 
We calculate the electromagnetic decay widths analytically, for the first time, without any further approximation. Specifically, our theoretical results for the $\Xi_c(2790)^{+/0}$ and $\Xi_c(2815)^{+/0}$ radiative decay widths, without the introduction of any additional parameters, display a significant agreement with the recent experimental values obtained by the Belle experiment. The agreement is due to the fact that we have not introduced further approximations to simplify the calculation of the difficult convective term. Our predictions may be useful for future experiments at the Belle, BABAR, and LHC experiments. 

\end{abstract}




\end{frontmatter}

\section{Introduction}
\label{introduction}
Over the last few decades,
more progress has been made in the experimental observation of radiative decays of singly charmed baryons~\cite{CLEO:1998wvk,BaBar:2006pve,Solovieva:2008fw,Belle:2020ozq}. In 1999, the CLEO collaboration~\cite{CLEO:1998wvk} observed the $\Xi'_c$ on using the electromagnetic decay to $\Xi_c \gamma$.
In 2006, the BABAR collaboration reported the first observation of the $\Omega^{*}_c$ using its radiative decay channel $\Omega^0_c\gamma$ \cite{BaBar:2006pve}. Later, in 2008, the Belle Collaboration reconstructed the $\Omega^{*}_c$ in the same electromagnetic channel \cite{Solovieva:2008fw}. However, because of the difficulty of the measurements, the electromagnetic decay widths of these ground states were not reported in Refs. ~\cite{CLEO:1998wvk,BaBar:2006pve,Solovieva:2008fw}.

More recently, in 2020, the Belle collaboration Ref.~\cite{Belle:2020ozq} used the branching ratios of the electromagnetic decays with respect to the total decay widths given in Ref.~\cite{Belle:2016lhy} in order to estimate the electromagnetic decay widths of the $P$-wave $\Xi_c(2790)^{+/0}$ and  $\Xi_c(2815)^{+/0}$ baryons. 
For the two neutral states, the resulting electromagnetic decay widths were large, namely $\Gamma(\Xi_c(2815)^0\rightarrow\Xi_c^0 \gamma) = 320 \pm 45^{+45}_{-80}$ KeV and $\Gamma(\Xi_c(2790)^0\rightarrow\Xi_c^0 \gamma) \sim 800 \pm 320$ KeV, while for charged states they obtained an upper limit given by $\Gamma(\Xi_c(2815)^+\rightarrow\Xi_c^+\gamma)< 80$ KeV and $\Gamma(\Xi_c(2790)^+\rightarrow\Xi_c^++\gamma) < 350$ KeV.
However, with the ongoing work of the Belle II experiment, it will soon be possible to measure other electromagnetic decay widths of excited single-charmed baryons, and with the high-luminosity upgrade of the Large Hadron Collider (HL-LHC).

On the theoretical side, the electromagnetic decays of singly charmed baryons have been studied in~\cite{Cheng:1992xi,Wang:2009ic,Wang:2009cd,Jiang:2015xqa,Dey:1994qi,Bernotas:2013eia,Aliev:2014bma,Aliev:2009jt,Aliev:2016xvq,Aliev:2011bm,Chow:1995nw,Bahtiyar:2016dom,Bahtiyar:2015sga,Ivanov:1998wj,Savage:1994wa,Banuls:1999br,Cho:1994vg,Wang:2018cre,Cheng:1997rp,Simonis:2018rld,Majethiya:2009vx,Hazra:2021lpa,Bijker:2020tns,Tawfiq:1999cf,Ivanov:1999bk,Wang:2017kfr,Gamermann:2010ga,Ortiz-Pacheco:2023kjn,Aliev:2019lfs,Peng:2024pyl}.
The works in Refs. ~\cite{Cheng:1992xi,Wang:2009ic,Wang:2009cd,Jiang:2015xqa,Dey:1994qi,Bernotas:2013eia,Aliev:2014bma,Aliev:2009jt,Aliev:2016xvq,Aliev:2011bm,Chow:1995nw,Bahtiyar:2016dom,Bahtiyar:2015sga,Ivanov:1998wj,Savage:1994wa,Banuls:1999br,Cho:1994vg,Wang:2018cre,Cheng:1997rp,Simonis:2018rld,Majethiya:2009vx,Hazra:2021lpa} studied 
the radiative transitions of $S$-wave singly charmed baryons. However, fewer studies have focused on the radiative decays of $P$-wave charmed baryons \cite{Bijker:2020tns,Tawfiq:1999cf,Ivanov:1999bk,Wang:2017kfr,Gamermann:2010ga,Ortiz-Pacheco:2023kjn,Aliev:2019lfs,Peng:2024pyl}, which are the subject of the present article.

In Ref.~\cite{Tawfiq:1999cf}, the authors applied heavy quark symmetry to the radiative decays of singly charmed baryons by taking the limit \(m_Q \rightarrow \infty\).  Nevertheless, they did not calculate the radiative transitions of \(\Xi_c\) baryons. In Ref.~\cite{Ivanov:1999bk}, the authors used an effective hadron-quark Lagrangian (EHQL) to calculate the radiative transitions of some charmed \(P\)-wave states. They only addressed the radiative transitions of \(\Xi_c(2815)^{+/0}\), although their results overestimated the experimental data.
The Nonrelativistic Quark Model (NRQM) has been used in Refs.~\cite{Bijker:2020tns,Wang:2017kfr,Ortiz-Pacheco:2023kjn,Peng:2024pyl} to study singly charmed baryons. Specifically, Refs.~\cite{Wang:2017kfr,Peng:2024pyl} apply the dimensional substitution introduced by Close and Copley~\cite{Close:1970kt}, and Refs.~\cite{Bijker:2020tns,Ortiz-Pacheco:2023kjn} adopt the substitutions $\mathbf{p}_{\lambda} \approx i m_{\lambda} k_0 \boldsymbol{\lambda}$ and $\mathbf{p}_{\rho} \approx i m_{\rho} k_0 \boldsymbol{\rho}$ in the convective term of the electromagnetic Hamiltonian.
Thus, while Refs.~\cite{Bijker:2020tns},~\cite{Wang:2017kfr},~\cite{Ortiz-Pacheco:2023kjn}, and~\cite{Peng:2024pyl} followed the experimental trends; they did not perform an exact evaluation of the convective term in the electromagnetic interaction Hamiltonian.
The Coupled-Channel Approach (CCA)~\cite{Gamermann:2010ga} has been utilized to investigate the radiative decays of charmed \(P\)-wave states, \(\Sigma_c(2792)^{\pm,0}\), \(\Xi_c(2790)^{+/0}\), \(\Xi_c(2970)^{+/0}\), and \(\Lambda_c(2595)\). Additionally, the QCD Sum Rules (QCDSR) approach has been applied to the study of the radiative decays of \(\Sigma_c(2792)^{\pm,0}\), \(\Xi_c(2790)^{+/0}\), \(\Xi_c(2870)^{+/0}\), and \(\Lambda_c(2592)\) ~\cite{Aliev:2019lfs}.
  The cases of  Refs.~\cite{Gamermann:2010ga} and~\cite{Aliev:2019lfs} followed the opposite trend for the $\Xi_c(2790)^0$ and $\Xi_c(2790)^+$ states. 

Here, we study the electromagnetic decays of \(P\)-wave \(\Xi_c\) baryons to ground states. We use the nonrelativistic limit of the Hamiltonian for the electromagnetic interaction while retaining all terms up to the order \(m_j^{-1}\). By utilizing an algebraic methodology based on ladder operators, we are able to analytically calculate the electromagnetic decay widths without any further approximations, in contrast with what was done in previous articles~\cite{Wang:2017kfr,Ortiz-Pacheco:2023kjn,Peng:2024pyl}.  The method used in this article is different from the derivative operator method used in Ref.~\cite{Garcia-Tecocoatzi:2023btk} in the bottom sector. Remarkably, our theoretical results for the \(\Xi_c(2790)^{+/0}\) and \(\Xi_c(2815)^{+/0}\) radiative decay widths are in good agreement with the recent experimental results by the Belle experiment \cite{Belle:2020ozq}.

\section{Electromagnetic decays}
\label{secEM}
The interaction Hamiltonian that describes the electromagnetic coupling between photons and quarks, at tree level, is given by
\begin{equation}
H = -\sum_j {\rm e}_j  {\bar q}_j  \gamma^{\mu} A_{\mu} q_j,
\end{equation}
where ${\rm e}_j$ and $q_j$  are the charge and the quark field, respectively, corresponding to the $j$-th quark, $\gamma^{\mu}$ are the Dirac matrices, and $A_{\mu}$ is the electromagnetic field. Taking the non-relativistic limit and keeping terms up to the order $m_j^{-1}$, the previous interaction leads to the following interaction Hamiltonian 
\begin{eqnarray}
&& \mathcal{H}_{\rm em} = \sum^3_{j=1}  \frac{1}{(2\pi)^{3/2}} \frac{{\rm e}_j}{(2{\rm k})^{1/2}}  \Bigg\{ \varepsilon^0 e^{-i\mathbf{k} \cdot \mathbf{ r}_j} \label{EqHamLeYaouanc} \\ 
&& - \frac{[ \mathbf{p}_j  \cdot \boldsymbol{\varepsilon} e^{-i\mathbf{k} \cdot \mathbf{ r}_j}  +  e^{-i\mathbf{k} \cdot \mathbf{ r}_j} \boldsymbol{\varepsilon} \cdot \mathbf{p}_j ] }{2m_j}- \frac{i\boldsymbol{\sigma} \cdot (\mathbf{k} \times \boldsymbol{\varepsilon}) e^{-i\mathbf{k} \cdot \mathbf{ r}_j} }{2m_j} \Bigg\} \nonumber, 
\end{eqnarray}
where $\mathbf{ r}_j$ and $\mathbf{ p}_j$, stand for the coordinate and momentum of the $j$-th quark, respectively, ${\rm k}$ is the photon energy and $\mathbf{ k}={\rm k} \hat{\mathbf z}$  corresponds to the momentum of a photon emitted in the positive $z$ direction. The polarization vector for radiative decays {is written as} $\varepsilon^{\mu}=(0,1,-i,0)/\sqrt{2}$. 
Notice that the zeroth component vanishes, $\varepsilon^0 = 0$
 , since radiative decays involve only real photons. Thus
 the first term of the Eq. (\ref{EqHamLeYaouanc}) is identically zero. On substituting the explicit form of the polarization vector, the interaction Hamiltonian describing the electromagnetic decays of baryons is found to be
\begin{equation}
\begin{split}
\mathcal{H}_{\rm em}&= \sqrt{\frac{4\pi}{(2\pi)^{3}{\rm k}}}\sum^3_{j=1}\mu_j\Big [{\rm k}\mathbf{ s}_{j,-}e^{-i\mathbf{k} \cdot \mathbf{ r}_j}-   \frac{1}{2}\left ( \mathbf{ p}_{j,-}e^{-i\mathbf{k} \cdot \mathbf{ r}_j}+e^{-i\mathbf{k} \cdot \mathbf{ r}_j}\mathbf{ p}_{j,-} \right) \Big],
\label{Hem}
\end{split}
\end{equation}
where  $\mu_j \equiv {\rm e}_j/(2m_j)$,  $\mathbf{ s}_{j,-} \equiv \mathbf{ s}_{j,x} - i \mathbf{ s}_{j,y}$, and $\mathbf{p}_{j,-}\equiv \mathbf{ p}_{j,x} - i \mathbf{ p}_{j,y}$, are the magnetic moment,  the spin ladder, and the momentum ladder operator of the $j$-th quark, respectively.  The Hamiltonian in Eq. (\ref{Hem}). consists of two parts. The first part is proportional to  the magnetic term which gives the  spin-flip transitions:
$
{\rm k}\mathbf{ s}_{j,-}e^{-i \mathbf{k} \cdot \mathbf{ r}_j}\equiv 
{\rm k}\mathbf{ s}_{j,-} \hat{U}_j $.
The second part of the Hamiltonian in Eq. (\ref{Hem}) is  the convective term, and gives the orbit-flip transitions:
$\mathbf{p}_{j,-} \, e^{-i \mathbf{k} \cdot \mathbf{ r}_j} + e^{-i \mathbf{k} \cdot \mathbf{ r}_j} \, \mathbf{p}_{j,-}\equiv \hat{T}_{j,-}$. By using the previous relations, the Hamiltonian of Eq. (\ref{Hem}) can be rewritten in terms of the $\hat{U}_j$ and $\hat{T}_{j,-}$ operators in the following compact form as in Ref.~\cite{Garcia-Tecocoatzi:2023btk} which was dedicated to singly bottom baryons:
\begin{eqnarray}
\mathcal{H}_{\rm em}=\sqrt{\frac{4\pi}{(2\pi)^{3}{\rm k}}}\sum^3_{j=1}\mu_j\Big [{\rm k}\mathbf{ s}_{j,-}\hat{U}_j-   \frac{1}{2}\hat{T}_{j,-} \Big].\label{Hem2}
\end{eqnarray}


It is important to point out that there are other approaches available in the literature to evaluate the radiative transitions. One of these was introduced in 1970 in the Appendix of Ref.~\cite{Close:1970kt}, in which Close and Copley used the well-known relation $i[H_0,\mathbf{r}_j]=\mathbf{p}_j/m_j$, with $H_0=\sum_{i}\mathbf{p}^2_i/(2m_i)$ free Hamiltonian of the constituent quarks, and the substitution of \mbox{$i[H_0,\mathbf{r}_j] $} with \mbox{$ i{\rm k}\mathbf{r}_j$}. Thus, they effectively rewrote the convective term of the Hamiltonian of Eq.(\ref{EqHamLeYaouanc}), by replacing $\mathbf{p}_j/m_j$ with $i{\rm k}\mathbf{r}_j$, as
\begin{equation}
\begin{split}
\mathcal{H}_{\rm conv} &= -\sum^3_{j=1}  \frac{1}{(2\pi)^{3/2}} \frac{{\rm e}_j}{(2{\rm k})^{1/2}}  \Bigg\{ \frac{\mathbf{p}_j  \cdot \boldsymbol{\varepsilon} e^{-i\mathbf{k} \cdot \mathbf{ r}_j}  +  e^{-i\mathbf{k} \cdot \mathbf{ r}_j} \boldsymbol{\varepsilon} \cdot \mathbf{p}_j }{2m_j} \Bigg\} \\
&\approx -i\sum^3_{j=1}  \frac{{\rm e}_j}{2(2\pi)^{3/2}}\sqrt{\frac{\rm k}{2}} \Bigg\{  \mathbf{r}_j  \cdot \boldsymbol{\varepsilon} e^{-i\mathbf{k} \cdot \mathbf{ r}_j}  +  e^{-i\mathbf{k} \cdot \mathbf{ r}_j} \boldsymbol{\varepsilon} \cdot \mathbf{r}_j   \Bigg\} \\
&\approx -i\sum^3_{j=1}  \frac{{\rm e}_j}{(2\pi)^{3/2}}  \sqrt{\frac{\rm k}{2}} \mathbf{r}_j  \cdot \boldsymbol{\varepsilon} e^{-i\mathbf{k} \cdot \mathbf{ r}_j} .
\label{conv} 
\end{split}
\end{equation}
Note that the substitution of \mbox{$i[H_0,\mathbf{r}_j] $} with \mbox{$ i{\rm k}\mathbf{r}_j$} cannot be derived rigorously, instead it follows from a dimensional analysis. Thus, the implementation of Eq.(\ref{conv}) in the study of electromagnetic transitions inherently involves the loss of valuable quantum information from the original system.
Iachello and Kusnesov also used this substitution in Ref.~\cite{Iachello:1992yu} to calculate radiative transitions for light mesons, advocating its use on account of the considerable simplification of the calculations. More recently, this approach has been widely implemented in electromagnetic studies of mesons and baryons. In particular,  it is used in the study of pseudoscalar meson photoproduction \cite{Li:1997gd}, in pion photoproduction ~\cite{Zhao:2002id}, electromagnetic transitions of bottomonium ~\cite{Deng:2016ktl}, 
radiative decays of singly heavy baryons ~\cite{Wang:2017kfr}, and radiative decays of singly charmed baryons ~\cite{Peng:2024pyl}.
In Ref.~\cite{Bijker:2020tns,Ortiz-Pacheco:2023kjn}, to calculate electromagnetic decays of heavy baryons, they replaced $\mathbf{ p}_{\lambda}$ with $i m_{\lambda} {\rm k_0} \boldsymbol{\lambda}$ and $\mathbf{ p}_{\rho}$ with $ i m_{\rho} {\rm k_0} \boldsymbol{\rho}$ (in an electromagnetic decay process $\rm k=k_0$) as confirmed by one of the authors \cite{Privatecomunication}. Unfortunately, this replacement was not declared in Ref.~\cite{Ortiz-Pacheco:2023kjn}. One can notice that this does not correspond to the replacement of $\mathbf{p}_j/m_j$ with $i{\rm k}\mathbf{r}_j$, as used by Close and Copley ~\cite{Close:1970kt}, previously discussed and accepted in the literature.
We emphasize that the adoption of any additional approximation is unnecessary, since the complete calculation can be performed analytically, as we carry out in the present study.

In the present study we also compare the electromagnetic decay widths of singly charmed baryons obtained directly from  Eq.(\ref{Hem}) with those calculated by using the approximation in Eq.(\ref{conv}) in the light of the experimental results presented in \cite{Belle:2020ozq}, and show that, even though the calculations are indeed simpler with the adoption of the Close and Copley's substitution, it should be used only as an estimate with respect to the full calculation using Eq.(\ref{Hem}), which can be calculated analytically in a straightforward manner, thereby providing results compatibles with the experimental values from Ref.~\cite{Belle:2020ozq}.

In \ref{AppendixEme}, we outline the procedure for obtaining the matrix elements for the  Hamiltonian of Eq. (\ref{Hem2}),

\begin{eqnarray}
A_{M_{J_A}}=\langle \phi_{A'},k_{A'},J_{A'}, {M_{J_A}}-1| \mathcal{H}_{\rm em} |\phi_A,k_{A},J_A, {M_{J_A}}\rangle,
\label{emtransition}
\end{eqnarray}
 where $ |\phi_{A},k_A,J_A, {M_{J_A}}\rangle$ and  $|\phi_{A'}k_{A'}, J_{A'}, {M_{J_A}}-1 \rangle $ are the initial and final states, respectively. These states are defined in \ref{AppendixEme} and they diagonalize the  Hamiltonian of Ref.~\cite{Garcia-Tecocoatzi:2022zrf,Santopinto:2018ljf}. This Hamiltonian is parametrized and fitted to the experimental masses of charmed baryons reported by the PDG \cite{Workman:2022ynf} (see Table I of Ref.~\cite{Garcia-Tecocoatzi:2022zrf}).
 In the present study,
the masses and assignments for the singly charmed baryons are taken from Ref.~\cite{Garcia-Tecocoatzi:2022zrf}.
The decay widths are calculated by using the following equation
\begin{eqnarray}
\Gamma_{\rm em}(A\rightarrow A'\gamma)=  \Phi_{A\rightarrow A'\gamma}\frac{4\pi}{2J_A+1}\sum_{{M_{J_A}}>0}
\left|A_{M_{J_A}} \right|^2,  \label{gammaEM}
\end{eqnarray}
where $J_A$ is the initial state total angular momentum, and $ \Phi_{A\rightarrow A'\gamma}$ is the phase space factor, which in the rest frame of the initial baryon is given by
$ \Phi_{A\rightarrow A'\gamma}=4\pi\,({E_{A'}}/{m_{A}})\,{\rm k}^2, $
with $E_{A'}=\sqrt{m^2_{A'}+{\rm k}^2}$ being the energy of the final state; $m_A$ and $m_{A'}$ are the masses of the initial and final baryon, respectively, and $
 {\rm k}=(m^2_{A}-m^2_{A'})/{2m_{A}}$ 
is the final-state photon energy. 

It must be emphasized that our model is highly predictive, as it needs no additional free parameters in order to fit the experimental results. Moreover, we present a new algebraic methodology, based on ladder operators, to calculate the electromagnetic decay widths in an exact analytical way (see \ref{AppendixEme}).


\begin{table*}[h!tp]

\begin{center}
\begingroup
\setlength{\tabcolsep}{1.75pt } 
\renewcommand{\arraystretch}{1.35} 

\begin{tabular}{c c c c  c  c  c c c  c  c c c} \hline 
$\mathcal{F}={\bf {\bar{3}}}_{\rm F}$  &    &    &    & $\Xi_{c}^{+} \gamma$  & $\Xi_{c}^{0} \gamma$  & $\Xi'^{+}_{c} \gamma$  & $\Xi'^{0}_{c} \gamma$  & $\Xi'^{*+}_{c} \gamma$  & $\Xi'^{*0}_{c} \gamma$ \\
$\Xi_c(snc)$  & $\mathbf{J^P}$  & $\vert l_{\lambda}, l_{\rho}, k_{\lambda}, k_{\rho} \rangle$  & $^{2S+1}L_{J}$  & KeV  & KeV  & KeV  & KeV  & KeV  & KeV  \\ 
\hline
$\Xi_c(2788)$  & $ \mathbf{\frac{1}{2}^-}$ & $\vert \,1\,,\,0\,,\,0\,,\,0 \,\rangle $ &$^{2}P_{1/2}$&$ 28 _{- 16 }^{+ 17 }$    &  $ 335 _{- 22 }^{+ 23 }$    &  $ 3.6 _{- 1.5 }^{+ 1.5 }$    &    $ 0.1 _{- 0.02 }^{+ 0.03 }$   &   $ 0.3 _{- 0.1 }^{+ 0.1 }$  &  0 \\
 & &  &  & $< 350$  & $\sim 800^{+320}_{-320}$  &  ...    &  ...   &  ...   &  ...    &  \cite{Belle:2020ozq} \\
 & &  &  & 4.65  & 263  &  1.43    &  0   &  0.44   &  0    &  \cite{Wang:2017kfr} \\
& &  &  & 5.4  & 239.3  &  2.3    &  0    &  0.2    &  0    &  \cite{Bijker:2020tns} \\
& &  &  & 1.7  & 217.5  &  1.2    &  0    &  0.5    &  0    &  \cite{Peng:2024pyl} \\
$\Xi_c(2815)$  & $ \mathbf{\frac{3}{2}^-}$ & $\vert \,1\,,\,0\,,\,0\,,\,0 \,\rangle $ &$^{2}P_{3/2}$&$ 20 _{- 13 }^{+ 15 }$    &  $ 380 _{- 23 }^{+ 24 }$ & $ 6 _{- 2.6 }^{+ 2.8 }$    & $ 0.1 _{- 0.03 }^{+ 0.03 }$    &    $ 0.7 _{- 0.3 }^{+ 0.3 }$   &  0 \\
 & &  &  & $< 80$  & $320^{+90}_{-125}$  &  ...    &  ...   &  ...   &  ...    &  \cite{Belle:2020ozq} \\
 & &  &  & 2.8  & 292  &  2.32   &  0  &  0.99  &  0  &  \cite{Wang:2017kfr} \\
& &  &  & 2.4  & 344.6  &  4.6    &  0.1    &  0.6    &  0    &  \cite{Bijker:2020tns} \\
& &  &  & 1  & 243.1  &  2.1    &  0    &  1.2    &  0    &  \cite{Peng:2024pyl} \\
$\Xi_c(2935)$  & $ \mathbf{\frac{1}{2}^-}$ & $\vert \,0\,,\,1\,,\,0\,,\,0 \,\rangle $ &$^{2}P_{1/2}$&$ 24 _{- 10 }^{+ 11 }$    &  $ 32 _{- 11 }^{+ 12 }$    &  $ 559 _{- 42 }^{+ 46 }$    &  $ 12 _{- 3 }^{+ 3 }$   &  $ 2.4 _{- 1 }^{+ 1 }$   &  $ 0.1 _{- 0.03 }^{+ 0.03 }$   \\
 & &  & & 1.39  & 5.57  &  128   &  0    &  0.25  &  0    &  \cite{Wang:2017kfr} \\
& &  &  & 16  & 26.2  &  157.2   &  3.3    &  1.8    &  0    &  \cite{Bijker:2020tns} \\
$\Xi_c(2977)$  & $ \mathbf{\frac{1}{2}^-}$ & $\vert \,0\,,\,1\,,\,0\,,\,0 \,\rangle $ &$^{4}P_{1/2}$&$ 16 _{- 7 }^{+ 8 }$    &  $ 22 _{- 8 }^{+ 9 }$    &  $ 8 _{- 4 }^{+ 4 }$    &  $ 0.2 _{- 0.07 }^{+ 0.07 }$    &  $ 86 _{- 17 }^{+ 15 }$    &  $ 2_{- 0.4 }^{+ 0.5 }$   \\
& &  &  & 0.75 & 3.0  &  0.41   &  0   &  43.4  &  0  &  \cite{Wang:2017kfr} \\
& &  &  & 9.8  & 16.1  &  5.4    &  0.1    &  12.5    &  0.3    &  \cite{Bijker:2020tns} \\
$\Xi_c(2962)$  & $ \mathbf{\frac{3}{2}^-}$ & $\vert \,0\,,\,1\,,\,0\,,\,0 \,\rangle $ &$^{2}P_{3/2}$&$ 29 _{- 11 }^{+ 13 }$    &  $ 39 _{- 12 }^{+ 14 }$    &  $ 1218 _{- 194 }^{+ 191 }$    & $ 26 _{- 9 }^{+ 10 }$    &  $ 3.4 _{- 1.4 }^{+ 1.5 }$   & $ 0.1 _{- 0.03 }^{+ 0.03 }$   \\
& &  &  & 1.88  & 7.5  &  110  &  0  &  0.52  &  0  & \cite{Wang:2017kfr} \\
& &  &  & 20.6  & 33.7  &  585.1    &  12.4    &  2.8    &  0.1    &  \cite{Bijker:2020tns} \\
$\Xi_c(3004)$  & $ \mathbf{\frac{3}{2}^-}$ & $\vert \,0\,,\,1\,,\,0\,,\,0 \,\rangle $ &$^{4}P_{3/2}$&$ 52 _{- 20 }^{+ 23 }$    &  $ 72 _{- 24 }^{+ 25 }$    &  $ 30 _{- 12 }^{+ 14 }$    & $ 0.6 _{- 0.2 }^{+ 0.3 }$    &  $ 415 _{- 32 }^{+ 34 }$   &  $ 9 _{- 2 }^{+ 3 }$   \\
 & &  &  & 2.81  & 11.2  &  1.85  &  0  &  58.1  &  0  & \cite{Wang:2017kfr} \\
& &  &  & 34.6  & 56.6  &  21.2    &  0.4    &  122.4    &  2.6    &  \cite{Bijker:2020tns} \\
$\Xi_c(3049)$  & $ \mathbf{\frac{5}{2}^-}$ & $\vert \,0\,,\,1\,,\,0\,,\,0 \,\rangle $ &$^{4}P_{5/2}$&$ 43 _{- 16 }^{+ 18 }$    &  $ 59 _{- 18 }^{+ 19 }$    &  $ 28 _{- 10 }^{+ 11 }$    & $ 0.6 _{- 0.2 }^{+ 0.3 }$    &  $ 1258 _{- 214 }^{+ 208 }$  &  $ 27 _{- 10 }^{+ 11 }$   \\
 & &  &  & ...  & ... & ... &  ...  & ... & ... &   \cite{Wang:2017kfr} \\
& &  &  & 30.8  & 50.5  &  21.7    &  0.5    &  445.5    &  9.5    &  \cite{Bijker:2020tns} \\
\hline 
\end{tabular}

\endgroup
\end{center}
\caption{Predicted $ \Xi_c(snc) $  electromagnetic decay widths (in KeV). The first column reports the baryon name with its predicted mass, as from Ref.~\cite{Garcia-Tecocoatzi:2022zrf}. The second column displays $\bf J^{\rm P}$. The third column shows the internal configuration of the baryon $\left| l_{\lambda},l_{\rho}, k_{\lambda},k_{\rho}\right\rangle$ within the three-quark model, where $l_{\lambda,\rho}$ represent the orbital angular momenta and $k_{\lambda,\rho}$ denote the number of nodes of the $\lambda$ and $\rho$ oscillators.
The fourth column presents the spectroscopic notation $^{2S+1}L_J$ for each state.  Furthermore, $N=n_\rho+n_\lambda=1$ is the energy band.  Starting from the fifth column, the electromagnetic decay widths, computed by means of Eq.~\ref{gammaEM}, are presented. Each column corresponds to an electromagnetic decay channel;  the decay products are indicated at the top of the column. The zero values are electromagnetic decay widths either too small to be shown on this scale or not permitted by phase space. Our results are compared with those of references \cite{Wang:2017kfr}, \cite{Bijker:2020tns}, and \cite{Peng:2024pyl}. The ``$...$" indicates that there is no prediction for that state in Ref. \cite{Wang:2017kfr}.
}
\label{cascadesEM}
\end{table*}


\begin{table*}
\begin{tabular}{c ccccccc c}
\hline
Decay channels  & NRQM~\cite{Bijker:2020tns} & RQM~\cite{Ivanov:1999bk} &  NRQM~\cite{Wang:2017kfr} & CCA~\cite{Gamermann:2010ga} &  QCDSR~\cite{Aliev:2019lfs} & GEM~\cite{Peng:2024pyl} & ours & Exp.~\cite{Belle:2020ozq} \\
\hline 
$\Gamma \left( \Xi_c(2790)^0 \rightarrow \Xi_c^0 \gamma \right)$   & 239.3 & ...  & 263 & 119.3 &  2.7 & 217.5 &  $ 335 _{- 22 }^{+ 23 }$ & $\sim 800 \pm 320$ \\
$\Gamma \left( \Xi_c(2790)^+ \rightarrow \Xi_c^+ \gamma \right)$   & 5.4 & ...  & 4.65 & 249.6 &  265 & 1.7 & $ 28 _{- 16 }^{+ 17 }$ &  $< 350$ \\
\hline 
$\Gamma \left( \Xi_c(2815)^0 \rightarrow \Xi_c^0 \gamma \right)$   & 344.6 &  497 & 292 & ... & ... & 243.1 & $ 380 _{- 23 }^{+ 24 }$ & $320 \pm 45^{+45}_{-80}$  \\
$\Gamma \left( \Xi_c(2815)^+ \rightarrow \Xi_c^+ \gamma \right)$   & 2.4 & 190 & 2.8 & ... &  ... & 1.0 & $ 20 _{- 13 }^{+ 15 }$ & $< 80$  \\
\hline
\end{tabular}
\caption{Comparison of our electromagnetic decay widths for the $\Xi_c(2790)^{+/0}$, $\Xi_c(2815)^{+/0}$ baryons (in KeV) with those of previous studies. The first column contains the decay channels. Our results are compared with those of references \cite{Bijker:2020tns} (second column), \cite{Ivanov:1999bk} (third column), \cite{Wang:2017kfr} (fourth column), \cite{Gamermann:2010ga} (fifth column), \cite{Aliev:2019lfs} (sixth column), and \cite{Peng:2024pyl} (seventh column). Our results are also compared with the experimental values from \cite{Belle:2020ozq}. The ``$...$" indicates that there is no prediction for that state in the references.
}
\label{EMcomparison}
\end{table*}


\section{Results and discussion}

Table~\ref{cascadesEM} presents our results for the electromagnetic decay widths of the $\Xi_c^{+/0}$ baryons in transitions from $P$-wave to ground states, compared to experimental data~\cite{Belle:2020ozq} and the results from Refs.~\cite{Wang:2017kfr,Bijker:2020tns,Peng:2024pyl}.
The decay widths are calculated analytically using Eqs.~(\ref{Hem2}), (\ref{emtransition}), and~(\ref{gammaEM}), with the associated uncertainties estimated using a Monte Carlo bootstrap method~\cite{Garcia-Tecocoatzi:2022zrf}.

 In the following, we discuss our theoretical results, which agree with the recent data reported by the Belle collaboration~\cite{Belle:2020ozq} for the $\Xi_c(2790)^{+/0}$ and $\Xi_c(2815)^{+/0}$ states. In that study, they 
used their branching ratio 
measurements, together with the measured total widths reported in Ref.~\cite{Belle:2016lhy}, to extract the partial electromagnetic decay widths.

{\it 3.1. $\Xi_c(2790)^0$}: For the $\Xi_c(2790)^0$ state, the Belle collaboration \cite{Belle:2020ozq} estimated  a branching ratio of $(7.9 \pm 2.0^{+1.7}_{-2.3}) \%$ 
of the total width corresponding to the 
electromagnetic decay. This implies a partial decay width of 
\begin{eqnarray}
\Gamma_{em}[\Xi_c(2790)^0 \rightarrow \Xi^0_c \gamma] \sim 800\pm 320~\rm{KeV}.
\end{eqnarray}
According to Ref.~\cite{Garcia-Tecocoatzi:2022zrf} this state corresponds to a $P_\lambda$-wave state with ${\bf J}^P = {\bf \frac{1}{2}}^-$. 
Our theoretical decay width for this state is $335^{+23}_{-22}$ KeV. Compared with the experimental value of $\sim800\pm320$ KeV reported in Ref.~\cite{Belle:2020ozq}, the theoretical result obtained here slightly underestimates the value reported by Belle. 

{\it 3.2. $\Xi_c(2790)^+$}: For the electromagnetic decay of the $\Xi_c(2790)^+$, the Belle collaboration~\cite{Belle:2020ozq} did not find any signal, but set an upper limit on the partial width, i.e.:
\begin{eqnarray}
\Gamma_{em}[\Xi_c(2790)^+ \rightarrow \Xi^+_c \gamma] < 350~\rm{KeV}.
\end{eqnarray}
As discussed in Ref.~\cite{Garcia-Tecocoatzi:2022zrf} this state corresponds to a $P_\lambda$-wave state with ${\bf J}^P = {\bf \frac{1}{2}}^-$. 
Our result for the electromagnetic decay width is $28^{+17}_{-16}$ KeV, which is compatible with the upper limit set by Ref.~\cite{Belle:2020ozq}.

{\it 3.3. $\Xi_c(2815)^0$}: Similarly, the Belle collaboration \cite{Belle:2020ozq} estimated the partial width of the $\Xi_c(2815)^0$ decaying to $\Xi_c^0\gamma$ to be
\begin{eqnarray}
\Gamma_{em}[\Xi_c(2815)^0 \rightarrow \Xi_c^0\gamma]= 320\pm 45^{+45}_{-80}~\rm{KeV}.
\end{eqnarray}
Ref.~\cite{Garcia-Tecocoatzi:2022zrf} assigns this state to a $P_\lambda$-wave state with ${\bf J}^P = {\bf \frac{3}{2}}^-$. 
Our result for the electromagnetic decay width for this state is $380^{+24}_{-23}$ KeV, which is in good agreement with the value reported in Ref.~\cite{Belle:2020ozq}.

{\it 3.4. $\Xi_c(2815)^+$}:
For the $\Xi_c(2815)^+$ state,  the Belle Collaboration \cite{Belle:2020ozq} also estimated the upper limit for the partial electromagnetic decay width of this state to the $\Xi^+_c$ as
\begin{eqnarray}
\Gamma_{em}[\Xi_c(2815)^+ \rightarrow \Xi_c^+\gamma] < 80~\rm{KeV},
\end{eqnarray}

In Ref.~\cite{Garcia-Tecocoatzi:2022zrf} this state is identified as a $P_\lambda$-wave state with ${\bf J}^P = {\bf \frac{3}{2}}^-$. 
Our theoretical decay width for this state is $20^{+15}_{-13}$ KeV, which is compatible with the upper limit set by Ref.~\cite{Belle:2020ozq}.



From Table~\ref{cascadesEM}, we observe that the deviations between our results and those of Refs.~\cite{Bijker:2020tns,Wang:2017kfr,Peng:2024pyl} are relatively small in some cases, while in others they reach up to one order of magnitude. These discrepancies arise mainly from differences in the treatment of the convective term and smaller contributions from the choice of wave functions. For instance, in the decay channel $\Xi_c(2977) \to \Xi_c^{+} \gamma$, our predicted decay width is 16 KeV, whereas the value reported in Ref.~\cite{Peng:2024pyl} is 0.75 KeV. 
In contrast, comparing Refs~\cite{Bijker:2020tns} and \cite{Wang:2017kfr}, the reasonable agreement observed for the $P_{\lambda}$-wave states can be attributed to the similarity in the substitutions employed. As seen in Table~\ref{cascadesEM}, their corresponding decay widths agree within deviations of approximately 10–40\%. However, for the $P_{\rho}$-wave states, their discrepancies are significantly larger, as shown in Table~\ref{cascadesEM}.

In addition, in Table~\ref{EMcomparison}, we compare our results for the decay widths of ($\Xi_c(2790)^{+/0} \rightarrow \Xi^{+/0}_c \gamma$) and ($\Xi_c(2815)^{+/0} \rightarrow \Xi_c^{+/0}\gamma$) with those theoretically obtained by means of the Non Relativistic Quark Model  (NRQM)~\cite{Bijker:2020tns,Wang:2017kfr},  Relativistic Quark Model (RQM)~\cite{Ivanov:1999bk}, Coupled-Channel Approach (CCA)~\cite{Gamermann:2010ga}, QCD Sum Rules (QCDSR)~\cite{Aliev:2019lfs}, Gaussian Expansion Method (GEM)~\cite{Peng:2024pyl}, and with the experimental values reported by the Belle collaboration~\cite{Belle:2020ozq}.

Table~\ref{EMcomparison} shows that the deviations between our results and those of Refs.~\cite{Bijker:2020tns,Wang:2017kfr} are approximately 10-30\% for the $\Xi_c(2815)^{0}$ and 27-40\% for the $\Xi_c(2790)^{0}$, both of which are compatible with the available experimental data. In contrast, for positively charged states, the deviations are significantly larger: around 700-800\% for the $\Xi_c(2815)^{+}$ and 500-600\% for the $\Xi_c(2790)^{+}$. In both Refs.~\cite{Bijker:2020tns} and~\cite{Wang:2017kfr}, the treatment of the convective term contributes to these differences. Our results show larger discrepancies when compared with those of Ref.~\cite{Peng:2024pyl}. This comparison is more complex, as Ref.~\cite{Peng:2024pyl} use the Close and Copley substitution and numerical wave functions. It is worth noting that Refs.~\cite{Bijker:2020tns,Wang:2017kfr} and the present work use harmonic oscillator wave functions. This observation suggests that, once more precise measurements for the positively charged $\Xi_c$ states become available, it will be possible to assess how much the dimensional substitutions used in Ref.~\cite{Bijker:2020tns} and in Ref.~\cite{Wang:2017kfr}, impact the accuracy of their predictions.

Moreover, as seen in Table~\ref{EMcomparison}, the electromagnetic amplitudes $A_{M_{J_A}}$ are sensitive to the choice of wave functions. This is evident from a comparison of the results of Refs.~\cite{Wang:2017kfr} and ~\cite{Peng:2024pyl}, which both use Close and Copley substitution to evaluate the electromagnetic transition amplitudes, and only differ since Ref.~\cite{Wang:2017kfr} uses harmonic oscillator wave functions, while Ref.~\cite{Peng:2024pyl} uses numerical wave functions. As shown in Tables~\ref{cascadesEM} and~\ref{EMcomparison}, the two results differ by up to a factor of about three. Comparing the results of the present paper with those of Ref. ~\cite{Wang:2017kfr,Peng:2024pyl}, they are always higher, and the differences can be as large as a factor of ten. Furthermore, if we compare our results with those of Ref. ~\cite{Bijker:2020tns} which uses the same harmonic oscillator wave functions as the present work, but adopts a dimensional substitution similar to that of Close and Copley to simplify the calculation of the electromagnetic widths, we see in Tables~\ref{cascadesEM} and~\ref{EMcomparison}, that the electromagnetic decay amplitudes obtained in the present work, are also in this case up to an order of magnitude larger than those of Ref.~\cite{Bijker:2020tns}. This larger discrepancy highlights the impact of using the full Hamiltonian of the electromagnetic interaction, without simplifications of the convective terms used in previous studies.
\section{Summary and conclusions}
\label{conclusions}

In this paper, we present our predictions for the electromagnetic decay widths of the $\Xi_c$ baryons from the $P$-wave states to the ground states by using the masses of Ref.~\cite{Garcia-Tecocoatzi:2022zrf} without introducing any further free parameters. We perform our calculations in a new analytical way as described in \ref{AppendixEme}.


Our results for the $\Xi_c(2790)^0$ and $\Xi_c(2815)^0$ states are in good agreement with the values reported in Ref.~\cite{Belle:2020ozq}. In the case of the $\Xi_c(2790)^+$ electromagnetic decay widths, the data are much less precise. Nevertheless, the experimental decay width is consistent with our theoretical value. Similarly, the $\Xi_c(2815)^+$ decay width is compatible with the upper limit set by the Belle Collaboration~\cite{Belle:2020ozq}.

The trend in the electromagnetic decay widths of the neutral and charged $\Xi_c(2815)$ and $\Xi_c(2790)$ states is well reproduced in our calculation, which is based on the assignment of the $\Xi_c(2790)$ and $\Xi_c(2815)$ as $P_\lambda$ states ~\cite{Garcia-Tecocoatzi:2022zrf}.


Our results demonstrate that the exact evaluation of the convective term can play an important role in the calculation of electromagnetic transitions that involve a change of angular momentum between the initial and final states, i.e. $\Delta L\neq  0$ transitions, as this term becomes dominant in such cases. 
The correct treatment of this term in the Hamiltonian ensures that our results are in good agreement with the available experimental data.

\section*{Acknowledgements}
A. R.-A. acknowledges support from the Secretar\'ia de Ciencia, Humanidades, Tecnolog\'ia e Innovaci\'on (Secihti), the Universidad de Guanajuato and INFN.  C.A. V.-A. is supported by the Secihti Investigadoras e Investigadores por M\'exico project 749 and SNI 58928.


\appendix

\section{Electromagnetic transition amplitudes}
\label{AppendixEme}
The transition amplitudes $A_{M_{J_A}} $ of Eq.~\ref{emtransition} read as,

\begin{eqnarray}
A_{M_{J_A}}=\langle \phi_{A'}, k_{A'},J_{A'}, {M_{J_A}}-1| \mathcal{H}_{\rm em} |\phi_A,k_{A},J_A, {M_{J_A}}\rangle.
\label{emtransition_App}
\end{eqnarray}
 Here $ |\phi_{A},k_A,J_A, {M_{J_A}}\rangle$ and  $|\phi_{A'},k_{A'}, J_{A'}, {M_{J_A}}-1 \rangle $ are the initial and final states, respectively.
  $\phi_A$ represents the flavor wave function, $k_A$ are the nodes, and
  $J_A$ is the total angular momentum with projection $M_{J_A}$. These charmed baryon states diagonalize the harmonic oscillator Hamiltonian of Ref.~\cite{Garcia-Tecocoatzi:2022zrf,Santopinto:2018ljf}.
 $\mathcal{H}_{\rm em}$ is the Hamiltonian of Eq. (\ref{Hem2})     given in terms of the operators $\hat{U}_j$ and $\hat{T}_{j,-}$.
To evaluate the transition amplitudes we write
\begin{eqnarray}
{A}_{M_{J_A}}&=& 2\sqrt{\frac{\pi}{{\rm k_0}}}{\rm k}\sum^3_{j=1} \Big [\langle \phi_{A'}, k_{A'},J_{A'}, {M_{J_A}}-1|\nonumber \\&\times& \hat{\mu}_j \mathbf{ s}_{j,-} \hat{U}_j |\phi_A,k_A,J_A, {M_{J_A}} \rangle \Big] \nonumber \\
&-& \sqrt{\frac{\pi}{{\rm k_0}}}\sum^3_{j=1}\Big [ \langle \phi_{A'}, k_{A'},J_{A'}, {M_{J_A}}-1| \nonumber \\&\times& \hat{\mu}_j \hat{T}_{j,-} |\phi_A, k_A,J_A, {M_{J_A}} \rangle\Big] .
\label{emtransition_sep}
\end{eqnarray}
Notice that the calculation is separable, since $\hat{\mu}_j$ acts in the flavor space, $\mathbf{s}_{j,-}$ operates in the spin space, and the  $\hat{U}_j$ and $\hat{T}_{j,-}$ operators act in the spatial space. Hence, we implement the following change of basis:
 \begin{eqnarray}
|\phi_A, k_A ,J_A,{M_{J_A}}\rangle &
=& 
\sum_{M_{L},M_{S}} \langle L, M_L;S, M_S|J_A,{M_{J_A}}\rangle \nonumber \\
& \times &  \sum_{m_{l_\lambda},m_{l_\rho}} \langle l_\lambda, m_{l_\lambda}; l_\rho, m_{l_\rho}|L,M_L\rangle \nonumber \\
& \times &  \sum_{m_{S_{12}},m_{S_3}} \langle S_{12},m_{S_{12}};S_3,m_{S_3}|S,M_S\rangle \nonumber \\
& \times & \sum_{m_{S_1},m_{S_2}}  \langle S_1, m_{S_1};S_2,m_{S_2} |S_{12},m_{S_{12}} \rangle \nonumber\\
& \times & |S_1,m_{S_1}\rangle \otimes | S_2,m_{S_2}\rangle \otimes | S_3,m_{S_3}\rangle\nonumber  \\
& \otimes &|\phi_A \rangle \otimes |k_\rho, l_\rho, m_{l_\rho}, k_\lambda, l_\lambda, m_{l_\lambda} \rangle ,
\end{eqnarray}
where $|S_i, m_{S_i}\rangle$ denotes the spin wave functions of each quark ($i=1,2,3$), and $|k_\rho, l_\rho, m_{l_\rho}, k_\lambda, l_\lambda, m_{l_\lambda} \rangle$ is  the spatial wave function, with $k_A = k_\rho + k_\lambda$.
The harmonic-oscillator spatial baryon wave function in momentum space is expressed in terms of $\omega_{\rho}$ and $\omega_{\lambda}$ by using the relation $\alpha^2_{\rho,\lambda}=\omega_{\rho,\lambda}m_{\rho,\lambda}$. We adopt the usual definitions for
$ n_{\rho(\lambda)}= 2 k_{\rho(\lambda)}+l_{\rho(\lambda)}$, with 
$k_{\rho(\lambda)}=0,1,...$, and $l_{\rho(\lambda)}=0,1,...$, where $l_{\rho(\lambda)}$ is the orbital angular momentum of the $\rho$($\lambda$) oscillator,  and $k_{\rho(\lambda)}$ is the number of nodes (radial excitations) in the $\rho$($\lambda$) oscillators. 
In the following, we use the notation \mbox{$\psi_{k_\rho,l_\rho,m_{l_\rho},k_\lambda,l_\lambda,m_{l_\lambda}}(\vec{p}_{\rho} ,\vec{p}_\lambda)=\langle \vec{p}_{\rho} ,\vec{p}_\lambda|k_\rho,l_\rho,m_{l_\rho},k_\lambda,l_\lambda,m_{l_\lambda}\rangle$.}

The matrix elements of the tensor operators $\hat{T}_{j,-}$, are  expressed as a sum of the matrix elements of the $\hat{U}_j$ operators. To achieve this, we calculate the action of the $\mathbf{p}_{\rho,\pm}$ and $\mathbf{p}_{\lambda,\pm}$ ladder operators on the states:
\begin{align}\label{EqT_U}
&\langle  k_{\rho_{A'}},l_{\rho_{A'}},m_{l_{\rho_{A'}}},k_{\lambda_{A'}},l_{\lambda_{A'}},m_{l_{\lambda_{A'}}}|\hat{T}_{j,-}|k_{\rho_A},l_{\rho_A},m_{l_{\rho_A}},k_{\lambda_A},l_{\lambda_A},m_{l_{\lambda_A}} \rangle \nonumber\\
&= \langle  k_{\rho_{A'}},l_{\rho_{A'}},m_{l_{\rho_{A'}}},k_{\lambda_{A'}},l_{\lambda_{A'}},m_{l_{\lambda_{A'}}}| \mathbf{p}_{j,-} \, \hat{U}_j |k_{\rho_A},l_{\rho_A},m_{l_{\rho_A}},k_{\lambda_A},l_{\lambda_A},m_{l_{\lambda_A}} \rangle \nonumber\\
&+\langle  k_{\rho_{A'}},l_{\rho_{A'}},m_{l_{\rho_{A'}}},k_{\lambda_{A'}},l_{\lambda_{A'}},m_{l_{\lambda_{A'}}}| \hat{U}_j \, \mathbf{p}_{j,-} |k_{\rho_A},l_{\rho_A},m_{l_{\rho_A}},k_{\lambda_A},l_{\lambda_A},m_{l_{\lambda_A}} \rangle.
\end{align}

\noindent
Then, the $\hat{T}_{j,-}$ is expressed as a sum of matrix elements of $\hat{U}_j$ weighted by the coefficients $C_\alpha$ and $C_{\beta}$
\begin{align}
&\langle  k_{\rho_{A'}},l_{\rho_{A'}},m_{l_{\rho_{A'}}},k_{\lambda_{A'}},l_{\lambda_{A'}},m_{l_{\lambda_{A'}}}|\hat{T}_{j,-}|k_{\rho_A},l_{\rho_A},m_{l_{\rho_A}},k_{\lambda_A},l_{\lambda_A},m_{l_{\lambda_A}} \rangle \nonumber\\ 
&= \sum_{\beta} C_\beta
\langle k_{\rho_{A'}},l_{\rho_{A'}},m_{l_{\rho_{A'}}},k_{\lambda_{A'}},l_{\lambda_{A'}},m_{l_{\lambda_{A'}}} | \nonumber \\
& \times \hat{U}_j  | k_{\rho_{A_\beta}},l_{\rho_{A_\beta}},m_{l_{\rho_{A_\beta}}},k_{\lambda_{A_\beta}}, l_{\lambda_{A_\beta}},m_{l_{\lambda_{A_\beta}}} \rangle \nonumber\\
&+ \sum_{\alpha} C_\alpha^{*}
\langle k_{\rho_{A'_\alpha}},l_{\rho_{A'_\alpha}},m_{l_{\rho_{A'_\alpha}}},k_{\lambda_{A'_\alpha}}, l_{\lambda_{A'_\alpha}},m_{l_{\lambda_{A'_\alpha}}} | \nonumber \\
& \times  \hat{U}_j  | k_{\rho_A},l_{\rho_A},m_{l_{\rho_A}},k_{\lambda_A}, l_{\lambda_A},m_{l_{\lambda_A}} \rangle.
\end{align}
In order to calculate the $C_\alpha$ and $C_{\beta}$ coefficients we write $
 \langle \vec{\rho}, \vec{\lambda}| \mathbf{p}_{j,-} |k_{\rho_A},l_{\rho_A},m_{l_{\rho_A}},k_{\lambda_A},l_{\lambda_A},m_{l_{\lambda_A}} \rangle
$ in terms of $
 \langle \vec{\rho}, \vec{\lambda}| \mathbf{p}_{\rho,-} |k_{\rho_A},l_{\rho_A},m_{l_{\rho_A}},k_{\lambda_A},l_{\lambda_A},m_{l_{\lambda_A}} \rangle
$ and $
 \langle \vec{\rho}, \vec{\lambda}| \mathbf{p}_{\lambda,-} |k_{\rho_A},l_{\rho_A},m_{l_{\rho_A}},k_{\lambda_A},l_{\lambda_A},m_{l_{\lambda_A}} \rangle
$ by using
\mbox{
$
 \mathbf{p}_{1,-} =  \frac{1}{\sqrt{2}}\mathbf{p_{\rho,-}}+\frac{1}{\sqrt{6}}\mathbf{p_{\lambda,-}}\,$,}
 \mbox{
 $
 \mathbf{p}_{2,-}  
= -\frac{1}{\sqrt{2}}\mathbf{p_{\rho,-}}+\frac{1}{\sqrt{6}}\mathbf{p_{\lambda,-}}\,$}, and $
\mathbf{p}_{3,-} =  -\sqrt{\frac{2}{3}} \mathbf{p_{\lambda, -}}\,
$.

In the following subsections, we evaluate the action of the ladder operators $\mathbf{p}_{\rho,\pm}$ and $\mathbf{p}_{\lambda,\pm}$ on the wave functions. This analysis allows us to identify the coefficients $C_{\alpha}$ and $C_{\beta}$.

\subsection{Ladder operators  in  momentum space}

In this work, we adopt a different strategy to evaluate the action of the ladder operators $\mathbf{p}_{\rho,\pm}$ and $\mathbf{p}_{\lambda,\pm}$, in contrast to the method used in Ref.~\cite{Garcia-Tecocoatzi:2023btk}, where the operators are treated as differential operators acting on coordinate-space wave functions. Here, we work in momentum space, where the ladder operators are represented as rank-1 irreducible tensor operators. This formulation enables a fully algebraic treatment: instead of calculating         derivatives, the operators act by transforming a given state into a linear combination of other states with well-defined angular momentum couplings. This significantly simplifies the evaluation of matrix elements and facilitates the direct implementation of $SU(2)$ algebra. The action of the ladder operators in this formalism is presented below.

\label{momentumlader}
In the momentum representation, the $\mathbf{p}_{j,\pm}$  ladder operators of the $j$-th quark are
\begin{eqnarray}
\mathbf{p}_{j,\pm}=\mp \sqrt{\frac{8\pi}{3}} |\mathbf{p}| Y^{\pm}_1(\mathbf{p})= \mp \sqrt{\frac{8\pi}{3}}\mathcal{Y}^{\pm1}_1 (\mathbf{p}), 
\end{eqnarray}
where $Y^{\pm}_1(\mathbf{p}) $ and 
 $\mathcal{Y}_{1}^{\pm}(\mathbf{p})$ are the rank-1 spherical and  solid harmonic, respectively. The above equation implies that, for the $\mathbf{p}_{\rho(\lambda),\pm} $ ladder operators, we have
\begin{eqnarray}
\mathbf{p}_{\rho(\lambda),\pm}&=& \mp \sqrt{\frac{8\pi}{3}}\mathcal{Y}^{\pm1}_1 (\mathbf{p}_{\rho(\lambda)}). 
\end{eqnarray} 


\noindent
The operators $\mathbf{p}_{\rho(\lambda),\pm}$ are diagonal in momentum space, that is,
\begin{eqnarray}
&& \langle \vec{p}_{\rho} ,\vec{p}_\lambda |\mathbf{p}_{\rho(\lambda),\pm} | {\vec{p}}^{\,\prime}_{\rho} ,{\vec{p}}^{\,\prime}_\lambda \rangle \nonumber \\
&& =  \mp \sqrt{\frac{8\pi}{3}}\mathcal{Y}^{\pm1}_1 (\vec p_{\rho(\lambda)}) \delta^3({\vec{p}}_{\rho}-{\vec{p}}^{\,\prime}_{\rho})
\delta^3({\vec{p}}_\lambda-{\vec{p}_\lambda}^{\,\prime}), 
\end{eqnarray}
\noindent
hence we are able to obtain, as an example, the action of the $\mathbf{p}_{\rho,\pm}$ operator on the ground state with 
\begin{eqnarray}
&& \langle {\vec{p}}_{\rho} ,{\vec{p}}_\lambda|\mathbf{p}_{\rho,\pm}|0,0,0,0,0,0\rangle \nonumber \\
&& = \int d^3 \vec{p}_{\rho}d^3 \vec{p}_{\lambda} \langle {\vec{p}}_{\rho} ,{\vec{p}}_\lambda|\mathbf{p}_{\rho,\pm} | {\vec{p}}^{\,\prime}_{\rho} ,{\vec{p}}^{\,\prime}_\lambda\rangle \langle {\vec{p}}^{\,\prime}_{\rho} ,{\vec{p}}^{\,\prime}_\lambda|0,0,0,0,0,0\rangle \nonumber \\
&& = \mp \sqrt{\frac{8\pi}{3}}\mathcal{Y}^{\pm1}_1({\vec{p}}_{\rho}) 3^{3/4}\;\left(\frac{1}{\pi
\omega_{\rho}m_{\rho}}\right)^{\frac{3}{4}}\,\left(\frac{1}{\pi
\omega_{\lambda}m_{\lambda}}\right)^{\frac{3}{4}} \nonumber \\
&& \times \exp\Big[{-\frac{
{\vec{p}}^{\,2}_{\rho}}{2\omega_{\rho}m_{\rho}} -\frac{
{\vec{p}}^{\,2}_{\lambda}}{2\omega_{\lambda}m_{\lambda}}}\Big]\nonumber \\ 
&& = \mp i  (\omega_{\rho}m_{\rho})^{1/2} \psi_{0,1,\pm 1,0,0,0}(\vec{p}_{\rho} ,\vec{p}_\lambda),
\label{pladderM}
\end{eqnarray}

The results for the other cases for the $\mathbf{p}_{\rho,\pm}$ operator are given by
\begin{eqnarray}
&& \langle {\vec{p}}_{\rho} ,{\vec{p}}_\lambda|\mathbf{p}_{\rho,\pm}|0,0,0,0,0,0\rangle \nonumber \\
&& = \mp i  (\omega_{\rho}m_{\rho})^{1/2}  \psi_{0,1,\pm 1,0,0,0}(\vec{p}_{\rho} ,\vec{p}_\lambda),\\
&& \langle {\vec{p}}_{\rho} ,{\vec{p}}_\lambda|\mathbf{p}_{\rho,\pm}|0,0,0,0,1,m_{l_\lambda}\rangle \nonumber \\
&& = \mp i  (\omega_{\rho}m_{\rho})^{1/2}  \psi_{0,1,\pm 1,0,1,m_{l_\lambda}}(\vec{p}_{\rho} ,\vec{p}_\lambda), \\
&& \langle {\vec{p}}_{\rho} ,{\vec{p}}_\lambda|\mathbf{p}_{\rho,-}|0,1,1,0,0,0\rangle \nonumber \\
&& =  \frac{i(\omega_{\rho}m_{\rho})^{1/2}}{\sqrt{3}} \psi_{0,2,0,0,0,0}(\vec{p}_{\rho} ,\vec{p}_\lambda)  \nonumber \\
&& + i  (\omega_{\rho}m_{\rho})^{1/2} \psi_{0,0,0,0,0,0}(\vec{p}_{\rho} ,\vec{p}_\lambda) \nonumber \\
&& + i\sqrt{\frac{2}{3}} (\omega_{\rho}m_{\rho})^{1/2} \psi_{1,0,0,0,0,0}(\vec{p}_{\rho} ,\vec{p}_\lambda), \\     
&& \langle {\vec{p}}_{\rho} ,{\vec{p}}_\lambda | \mathbf{p}_{\rho,-}|0,1,-1,0,0,0\rangle \nonumber \\
&& =  i\sqrt{2} (\omega_{\rho}m_{\rho})^{1/2} \psi_{0,2,-2,0,0,0}({\vec{p}}_{\rho} ,{\vec{p}}_\lambda) ,\\
&& \langle {\vec{p}}_{\rho} ,{\vec{p}}_\lambda | \mathbf{p}_{\rho,-}|0,1,0,0,0,0\rangle \nonumber \\
&& = i (\omega_{\rho}m_{\rho})^{1/2} \psi_{0,2,-1,0,0,0}({\vec{p}}_{\rho} ,{\vec{p}}_\lambda).
\end{eqnarray}

With analogous results for the action of the $\mathbf{p}_{\lambda,\pm}$ operator. 
Having calculated the action of the $\mathbf{p}_{\rho,\pm}$ and $\mathbf{p}_{\lambda,\pm}$ operators on the states, the $\hat{T}_{j,-}$ matrix elements are expressed as matrix elements of the $\hat{U}_{j}$ operator according to eq.\ref{EqT_U}.


\end{document}